\documentclass{article}
\usepackage[T1]{fontenc}
\usepackage[utf8]{inputenc}
\usepackage{graphics}
\usepackage{graphicx}
\usepackage[margin=2cm]{geometry}
\usepackage{caption}
\usepackage{subcaption}
\usepackage{color}
\usepackage{natbib}
\usepackage{amsmath}
\usepackage{float}

\title{A Systematic Comparison of q-Gaussian Fitting Methods across Complex Systems}
\author{\small{Facundo Abaca$^{1}$, Dario Javier Zamora$^{1}$\thanks{E-mail: djzamora@conicet.gov.ar}, Bruno S. Zossi$^{1}$, Ana G. Elias$^{1}$}, \\
\small{$^1$ Instituto de Fisica del Noroeste Argentino, CONICET and Universidad Nacional de Tucum\'an}\\
\small{ Av. Independencia 1800, Tucuman, CP 4000, Argentina}}

\date{\today}

\begin{document}

\maketitle

\begin{abstract}
Nature exhibits a wide variety of nonlinear phenomena characterized by strong fluctuations that push the system away from equilibrium. Recent studies have shown that non-extensive statistics are more suitable for analyzing these phenomena, as they can capture long-range correlations and heavy-tailed distributions that conventional Boltzmann-Gibbs statistics cannot represent. However, estimating the non-extensiveness parameter ($q$) is not trivial, and methods vary significantly across disciplines.

This paper presents a systematic and interdisciplinary comparison of three fitting methods: direct nonlinear fitting of the probability distribution function (PDF), linearization using the q-logarithm function, and numerical fitting of the cumulative distribution function (CDF). We applied this analysis to various phenomena, from areas such as geophysics, space weather and economics, analyzing the critical impact of discretization. The results show that $q$ estimates are highly sensitive to histogram binning, while the CDF-based method provides a bin-free alternative that yields more stable estimates for the datasets analyzed here. This contribution provides relevant information for the search for a method that allows the correct determination of the parameters of heavy-tailed distributions in various disciplines.

\end{abstract}

\section{Introduction}
Boltzmann-Gibbs statistics represents the conventional statistical framework applied to the majority of physical systems; however, it is fundamentally grounded in the assumptions of ergodicity and thermodynamic equilibrium. Nevertheless, these suppositions may fail in systems characterized by chaotic dynamics, sensitivity to initial conditions, memory effects, or long-range interactions \citep{Umarov2008}. These systems exist and are more common than one might think, given that nature often presents feedback mechanisms and self-similar structures that prevent the system from reaching molecular chaos or simple thermal equilibrium. These behaviors are relevant in disciplines such as astrophysics, geophysics, social and economic systems, and biology \citep{Carvalho2009,livadiotisUnderstandingKappaDistributions2013a, https://doi.org/10.1029/2021WR030506, Megias2023}.

The inhomogeneous transfer of a quantity leads to its concentration in localized spatial regions, producing intermittent and high fluctuations. As a consequence, the probability distribution functions (PDFs) develop enhanced tails populated by these events, resulting in heavy-tailed distributions \citep{Burlaga2004,Burlaga2005}.  In this regime, higher-order statistical moments become especially relevant, particularly when fluctuations are comparable to or exceed the mean value. To capture this different behavior, Constantino Tsallis proposed in 1988 the framework of non-extensive statistical mechanics \citep{Tsallis1988,Tsallis2005a}. By extremizing the nonadditive entropy $S_q=\sum (p_i^q-1)/(1-q)$, one derives the q-Gaussian distributions

\begin{equation}
    Y(x)= a(1-(1-q)b x^2)^{1/(1-q)}
\end{equation}

  This theoretical framework introduces the entropic index $q$, which quantifies the degree of departure from Gaussian statistics within a system. $q$ varies according to the dynamic properties of the system; when $q=1$ we return to the standard normal distribution, and for values $q>1$ we are dealing with heavy-tailed distributions, which are more common in systems out of equilibrium.

Despite the theoretical success of Tsallis statistics in describing nonlinear phenomena, there is no general consensus method for estimating the entropic index q. In the literature, various disciplines employ disparate fitting procedures, ranging from non-linear regression on PDFs to linearized methods using the q-logarithm function. However, these traditional approaches often suffer from the binning problem where the arbitrary choice of histogram intervals can lead to biased results and loss of information in the noisy tails \citep{nair2022fundamentals}.

This work addresses this methodological problem by presenting a comparison of fitting techniques including a numerical fitting of the CDF, applied to diverse complex systems as solar wind proton density, bitcoin closing value and river outflow. We aim to demonstrate that the choice of methodology is not merely a technical detail, but a critical factor in correctly capturing the heavy-tailed behavior
 of a system. We do not seek to establish a unified physical mechanism among these systems, but rather to test the stability of different q-Gaussian fitting procedures on datasets with distinct origins, sizes, and fluctuation structures.

This article is organized as follows: Section 2 presents a description of the set of complex systems analyzed in this study. Section 3 details the three fitting methodologies used to estimate the entropic q. Section 4 presents our main results. Finally, Section 5 summarizes our conclusions, analyzing the implications of our findings for the study of nonlinear systems. 

\section{Studied Systems}
We examine the proton density fluctuations of the solar wind using hourly averages from the OMNI directory covering the last 45 years. To ensure interdisciplinary rigor. We analyze the price variations of cryptocurrencies, choosing Bitcoin because it has a high market capitalization. In addition, we study the daily flow of the Brazos River, useful information for the design and management of water resource. The three systems are a classic example of heavy-tail behavior in natural phenomena
\subsection{Solar wind proton density}
The Sun continuously expels plasma because charged particles can gain sufficient energy to escape the Sun's gravitational pull; this is known as the solar wind. This low-density, high-speed stream of charged particles serves as a vast natural laboratory for the study of intense turbulence and nonlinear field fluctuations. In this medium, energy is transferred from large scales to smaller ones through a turbulent cascade. At sufficiently small scales, dissipative and dispersive processes transform this energy into other forms, such as thermal energy or particle acceleration. Scale invariance and self-similarity lead to fluctuations that typically exhibit power-law scaling, which causes energy to concentrate in localized regions and populates the tails of PDFs with highly energetic events, resulting in the characteristic heavy-tailed distributions\citep{Burlaga2009,Zamora2025,zhaoTurbulenceWavesTaylors2024}.

For this study, we utilize hourly mean values of the solar wind proton density ($N_p$) measured at the L1 Lagrange point. Figure 1 shows Np hourly values along the period. The data is retrieved from the NASA OMNI directory \citep{King2005}, which compiles near-Earth interplanetary magnetic field and plasma parameters from various spacecraft, primarily Wind and ACE.

\begin{figure}[h]
    \centering
    \includegraphics[width=\linewidth]{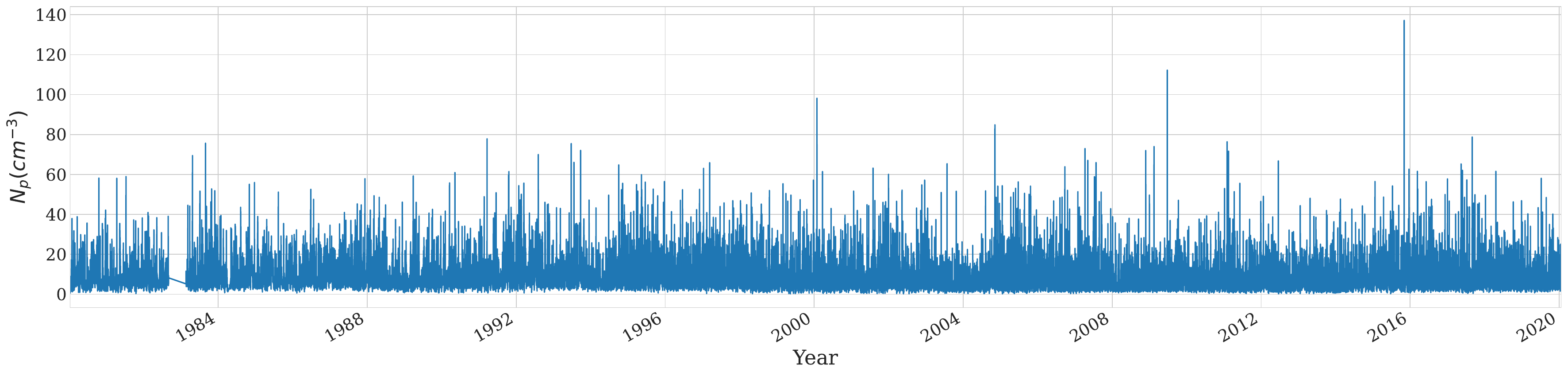}
    \caption{Values of proton density ($N_p$), in the solar wind for 1980 to 2025.}
    \label{fig:placeholdersolarwind}
\end{figure}

The variable analyzed consists of the successive fluctuations in proton density, defined as:
\begin{equation}
    dN_p(i)=2*\frac{N_p(i+1)-N_p(i)}{N_p(i+1)+N_p(i)}
\end{equation}
These fluctuations are adequately normalized using a local moving average to ensure statistical consistency.

\subsection{Closing prices of Bitcoin}
Cryptocurrencies represent a novel class of financial assets characterized by extreme price fluctuations, absence of centralized regulation, and high sensitivity to global news, making them ideal candidates for non-extensive statistical analysis. Unlike traditional stock markets, Bitcoin operates 24/7, exhibiting intermittent bursts of volatility and heavy-tailed distributions of returns that defy the standard Efficient Market Hypothesis\citep{stosicNonextensiveTripletsCryptocurrency2018, PhysRevE.104.054140}. In this study, we analyze the daily closing prices of Bitcoin (BTC) from 2016 to 2026, shown in Figure 2, obtained from the Web site https://coinmarketcap.com/currencies/ by market capitalization. 

\begin{figure}[h]
    \centering
    \includegraphics[width=\linewidth]{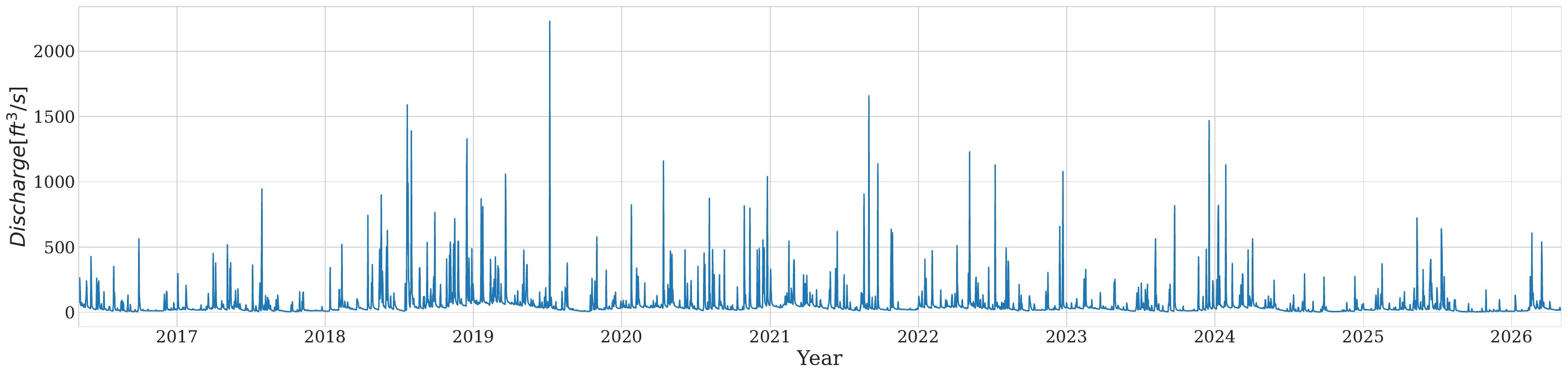}
    \caption{Closing values of bitcoin from 2016 to 2026.}
    \label{fig:Closing values of bitcoin from 2016 to 2026}
\end{figure}

For all the data in the time series we calculated the corresponding returns or logarithmic returns $r_t(i)$, in order to cancel systematic inflation and obtain a fluctuation around zero \citep{e19090457}.
\begin{equation}
    r_t=\ln\frac{P_b(i+1)}{P_b(i)}
\end{equation}

\subsection{Brazos River discharge}
Hydrological processes, such as river streamflow, are governed by complex interactions between precipitation, evapotranspiration, and basin characteristics, resulting in highly nonlinear and non-stationary dynamical systems. River discharge series often exhibit intermittent bursts, long-range temporal dependencies, and extreme events that lead to heavy-tailed probability distributions \citep{stosicTripletBrazosRiver2018,MIHAILOVIC2019290}. In this study, we analyze daily discharge data from the Brazos river in Texas, USA, for a 10-year period spanning from 3 May 2016, to 3 May 2026, as shown in Figure 3. The data is retrieved from the United States Geological Survey (USGS) National Water information Systems from the website https://waterdata.usgs.gov/monitoring-location/USGS-08110200. We proceed following the same methodology for complex time series used in the solar wind proton density time series.
\begin{figure}[h]
    \centering
    \includegraphics[width=\linewidth]{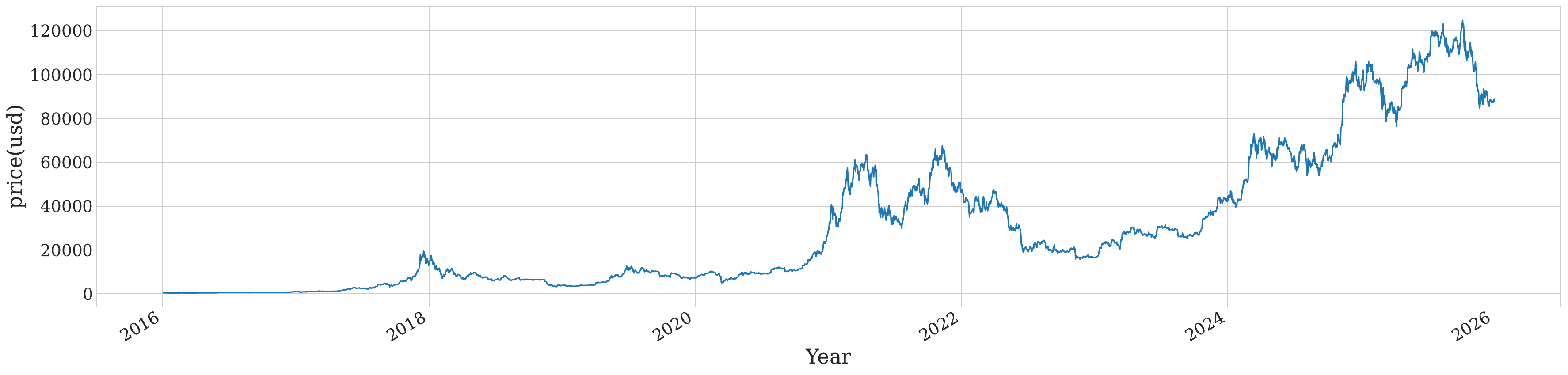}
    \caption{Discharge values of Brazos river for 2016 to 2026.}
    \label{fig:Variación temporal de la descarga del rio}
\end{figure}
\section{Methods}
In this section we present in details the three fitting methodologies used to estimate the entropic q: (i) direct nonlinear fitting to the PDF, (ii) the linearized method using the q-logarithm function (ln q) to identify the best correlation in the transformed space, and (iii) numerical fitting of the CDF.
\subsection{Direct non-linear fitting to the PDF}
To perform this fitting, a histogram is first constructed from the values of the analyzed variables, organizing the data into equal-width windows. This process requires selecting the number of bins, a choice that often relies on subjective criteria and can significantly influence the resulting parameters. Although heuristic rules such as 'Sturges' method are frequently employed to ensure algorithmic convergence, this method tends to underestimate the number of bins, and therefore increases noise in the tails of the distribution.

Following the binning process, a series of data points is generated using the center and the normalized height (relative frequency) of each bin. Finally, the q-Gaussian distribution, presented in Eq(1), is fitted to these points using a non-linear regression method. From this fit, the parameters of interest are extracted; specifically \textbf{$q_{nl}$}, which characterizes the departure from Gaussian statistics of the system, the normalization constant \textbf{a}, and the scale parameter \textbf{$b$}, which is intimately related to the width and the standard deviation of the PDF.
\subsection{Linear fitting to the PDF}
This second method is widely used due to its visual clarity; it relies on algebraic transformations to linearize the q-Gaussian distribution. We begin by defining the q-logarithm function.
\begin{equation}
    ln_qY=\frac{Y^{1-q}-1}{1-q}
\end{equation}
Applying this transformation to the q-Gaussian probability density:
\begin{equation}
    ln_qY=\frac{\left[a\cdot(1-(1-q)bx^2)^{1/1-q}\right]^{1-q}-1}{1-q}
\end{equation}
\begin{equation}
    ln_qY=\frac{a^{1-q}-1}{1-q}-a^{1-q}bx^2=A+Bx^2
\end{equation}
Where $A=ln_qa$ is the y-intercept and $B=-a^{1-q}b$ represents the slope of the linear relationship between the q-logarithm of the probability and the square of the normalized fluctuations

Following the histogram-based approach, we organize the data into bins, a process that inherently involves a degree of subjectivity in selecting the bin width, which can impact the stability of the results. Using the bin centers and heights, we process the data to test the linearized model: center values are squared ($x^2$), and q-logarithms are applied to the heights. Since the $ln_q$ function requires the value of $q$, the parameter we aim to determine, we implement an iterative search. We propose a range of trial values goin for 1 to 3 in steps of $0.01$. For each step, a linear regression is performed, and we select the $q_l$ that yields the highest coefficient of determination.

\subsection{Fitting of the CDF}
To overcome the limitations and subjective biases introduced by the binning process in traditional histograms, we implement a fitting method based on the Cumulative Distribution Function (CDF). This approach is particularly robust for heavy-tailed distributions, as it utilizes the full information of the dataset without loss of resolution in the noisy tails

The CDF, denoted as $F(x)$, is defined as:

\begin{equation}
    F(x)=P[X\leq x]
\end{equation}

The CDF evaluated at any number x is the probability that the random variable takes a value less than or equal to x. F can be obtained from the integral of the probability density function (PDF) from the lower limit to a point $x$.
\begin{equation}
    F(x)=\int_{-\infty}^{x}p(x')dx'
\end{equation}
Where $p(x')$ is the q-Gaussian distribution defined in Eq(1), since q-Gaussian distributions do not possess a simple closed-form integral in terms of elementary functions, we seek an analytical solution through the Beta function. By applying a change of variables,$\mu=x^2$ and $\nu=\frac{1}{q-1}$, the integral can be expressed as:
\begin{equation}
    \int_0^{\infty}\frac{\mu^{-1/2}}{[a(1+b/\nu \mu)]^{\nu}}d\mu=\left(\frac{b}{\nu}\right)^{-1/2}B\left[\frac{1}{2},\left(\nu-\frac{1}{2}\right)\right]=\left(b\right)^{-1/2}B\left[\frac{1}{2},-\left(\frac{1}{1-q}+\frac{1}{2}\right)\right]
\end{equation}
Where $B(x,y)$ is the Beta function, related to the Gamma function ($\Gamma$) by the identity:
\begin{equation}
    B(x,y)=\frac{\Gamma(x)\Gamma(y)}{\Gamma(x+y)}
\end{equation}
The Gamma function serves as the fundamental normalization tool for q distributions. 

For the implementation of this method, we perform a non-linear fit of the empirical CDF against the numerically integrated theoretical curve to minimize the residuals of the parameters $q$ and $b$. We obtain the empirical cumulative distribution function by arranging the data in ascending order and summing them until we have them all.

\section{Results}
In this section we presents our main results, offering a systematic comparison of the performance of each method across all analyzed systems. We analyze the stability of the parameter q and how its estimation is influenced by the number of intervals in traditional histograms, as well as the results obtained using the CDF-based approach.

The first analysis we performed with the datasets was to verify that they indeed exhibited non-gaussian behavior. If this is the case, we expect to find a greater number of large fluctuations in the histograms, which would be best fitted with a heavy-tailed distribution function, such as the q-Gaussian. In Figures 4, 5, and 6, we have the event density versus the fluctuations on a log scale at the "y" axes. We can see the height and center values of the bins for the various systems for the complete dataset, represented by the points. We selected a total of 100 bins as an example; we did not use a specific criterion to choose this number. Later, we will discuss some factors to consider when choosing this value for each data set. The figures also show the fitted q-Gaussian and a Gaussian distribution function.

\begin{figure}[H]
    \centering
    \includegraphics[width=0.75\linewidth]{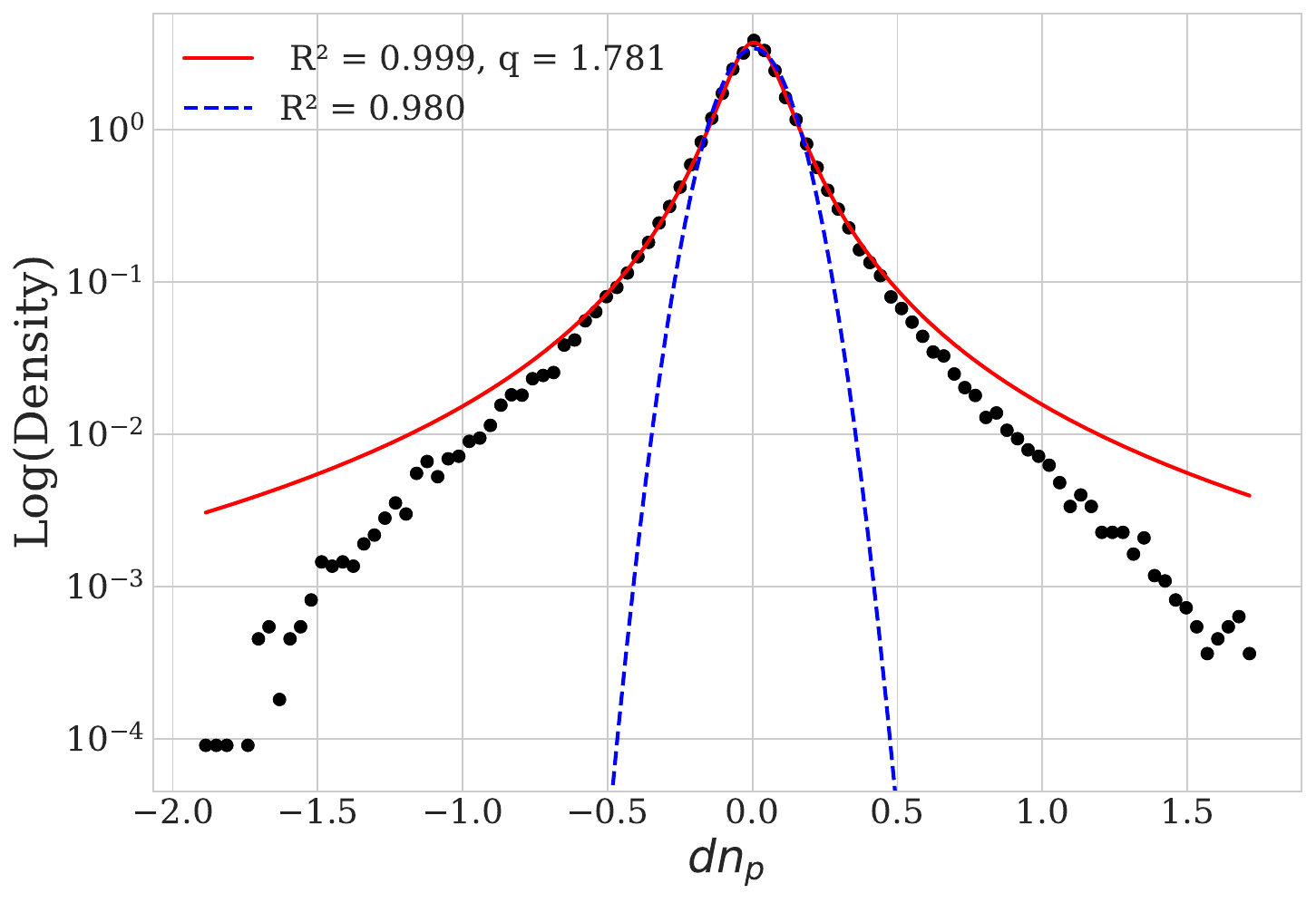}
    \caption{Comparison of the fit of the solar wind fluctuations histogram on a LOG scale.The red line is for the q-Gaussian distribution function and the blue one for the Gaussian distribution function.  }
    \label{fig:placeholder1}
\end{figure}

\begin{figure}[H]
    \centering
    \includegraphics[width=0.75\linewidth]{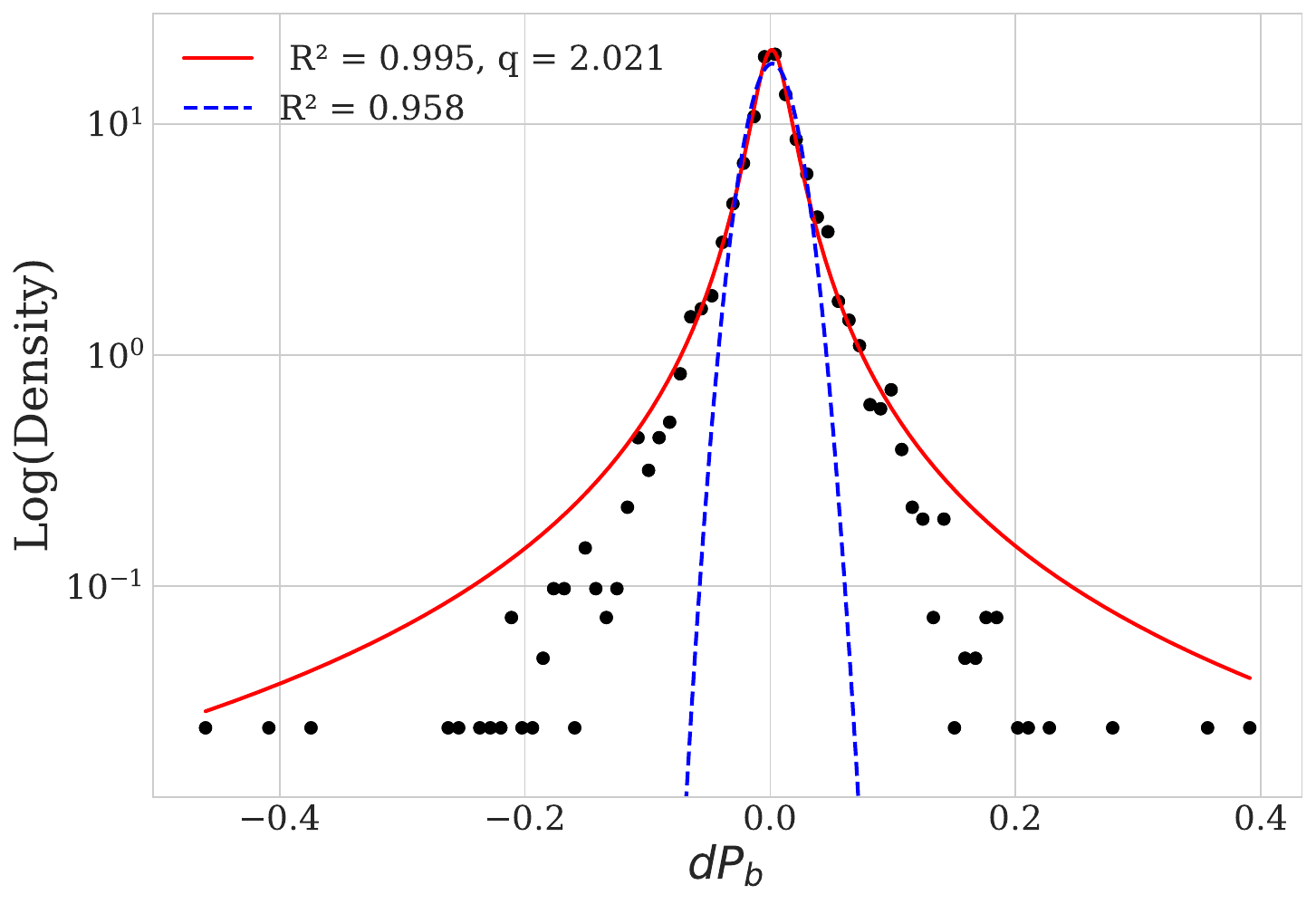}
    \caption{Comparison of the fit of logarithmic returns of the bitcoin prices fluctuations histogram on a LOG scale.}
    \label{fig:placeholder2}
\end{figure}

\begin{figure}[H]
    \centering
    \includegraphics[width=0.75\linewidth]{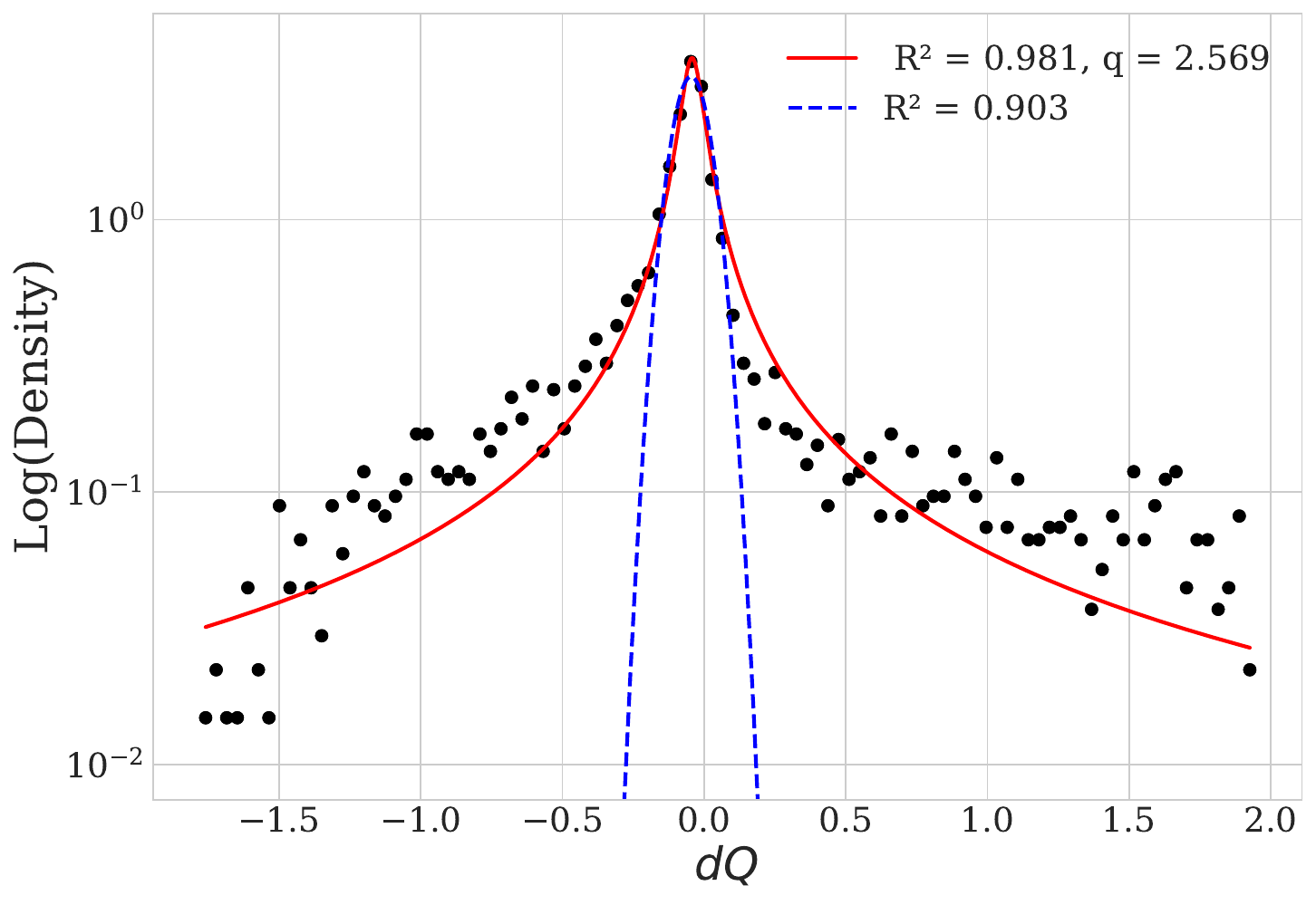}
    \caption{Comparison of the fit of the Brazos river discharge fluctuations histogram on a LOG scale.}
    \label{fig:placeholder3}
\end{figure}

A better fit was observed with heavy-tailed distributions, which was particularly noticeable on the logarithmic scale, especially at extreme values. Once it was confirmed that the system fit these distributions better, we proceeded to determine the sensitivity of the parameter q to the number of selected bins. To do this, we performed the fit on the three datasets with different bins values.

Starting with the direct fit, we proceeded working with an interval of 10 to 200 bins, in steps of 2, for the three cases here analyzed. The q values obtained in each step are shown in Figures 7, 8, and 9 (black dots). For the linear method, the q value was also estimated for different bins ranging from 10 to 200. They are aksi shown in Figures 7 to 9 (blue dots). It is important to remember that to proceed with the linearization of the data, it is first necessary to propose a value for q, which is modified until the best fit is found. Therefore, the error in the determination is equal to the value of the step used, 0.01 in our case. 

For the method based on the CDF, which do not depend on bins, a unique q value is obtained. This value is shown, together with values obtained with the other methods, in Figures 7 yo 9 (red dashed horizontal line). 


\begin{figure}[H]
    \centering
    \includegraphics[width=0.75\linewidth]{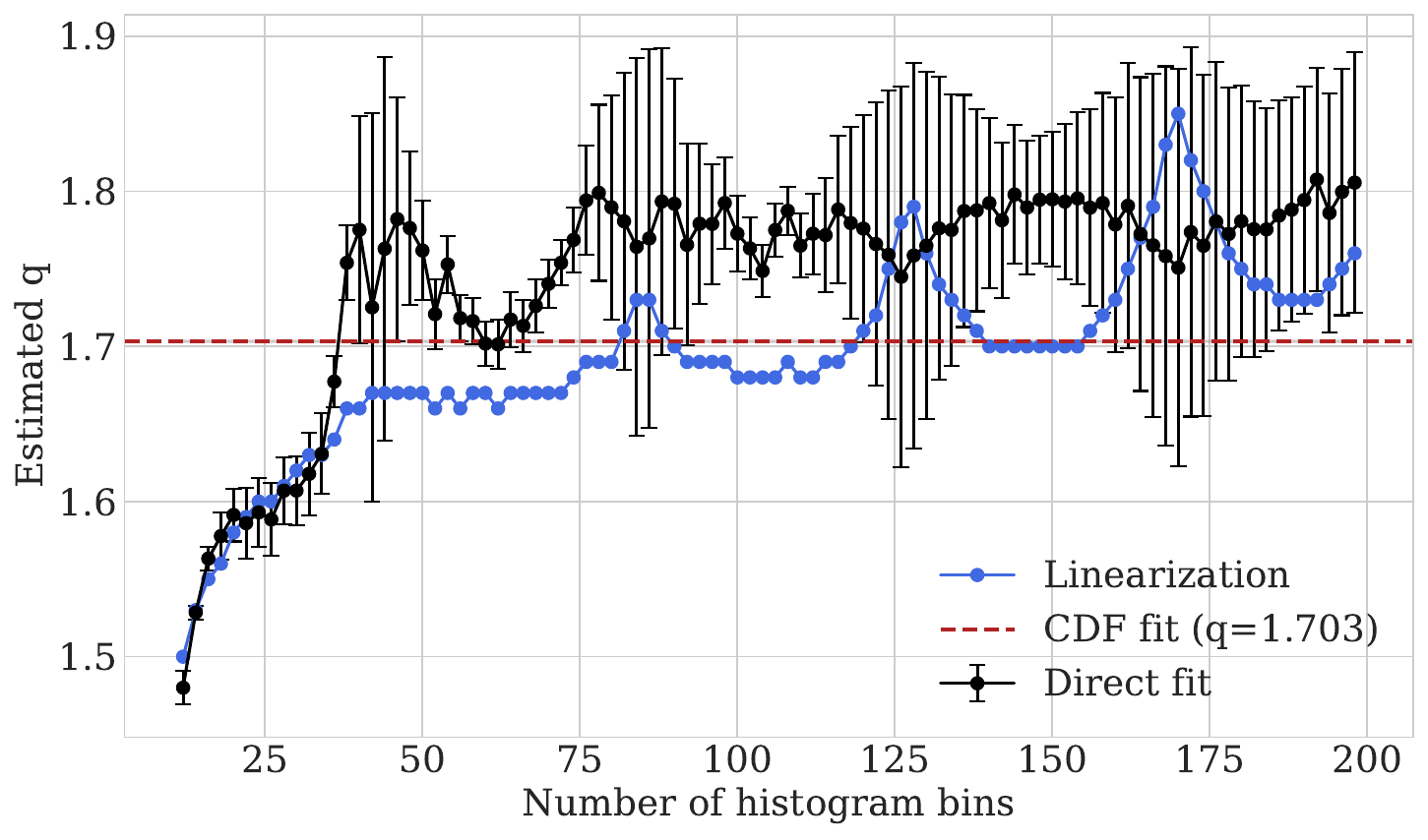}
    \caption{The parameter q, for the proton density distribution, in terms of the number of bins obtained with: the direct fit (black dots), the linearaized fit (blue dots), and CDF (red dashed line).}
    \label{fig:placeholderwadvd}
\end{figure}

\begin{figure}[H]
    \centering
    \includegraphics[width=0.75\linewidth]{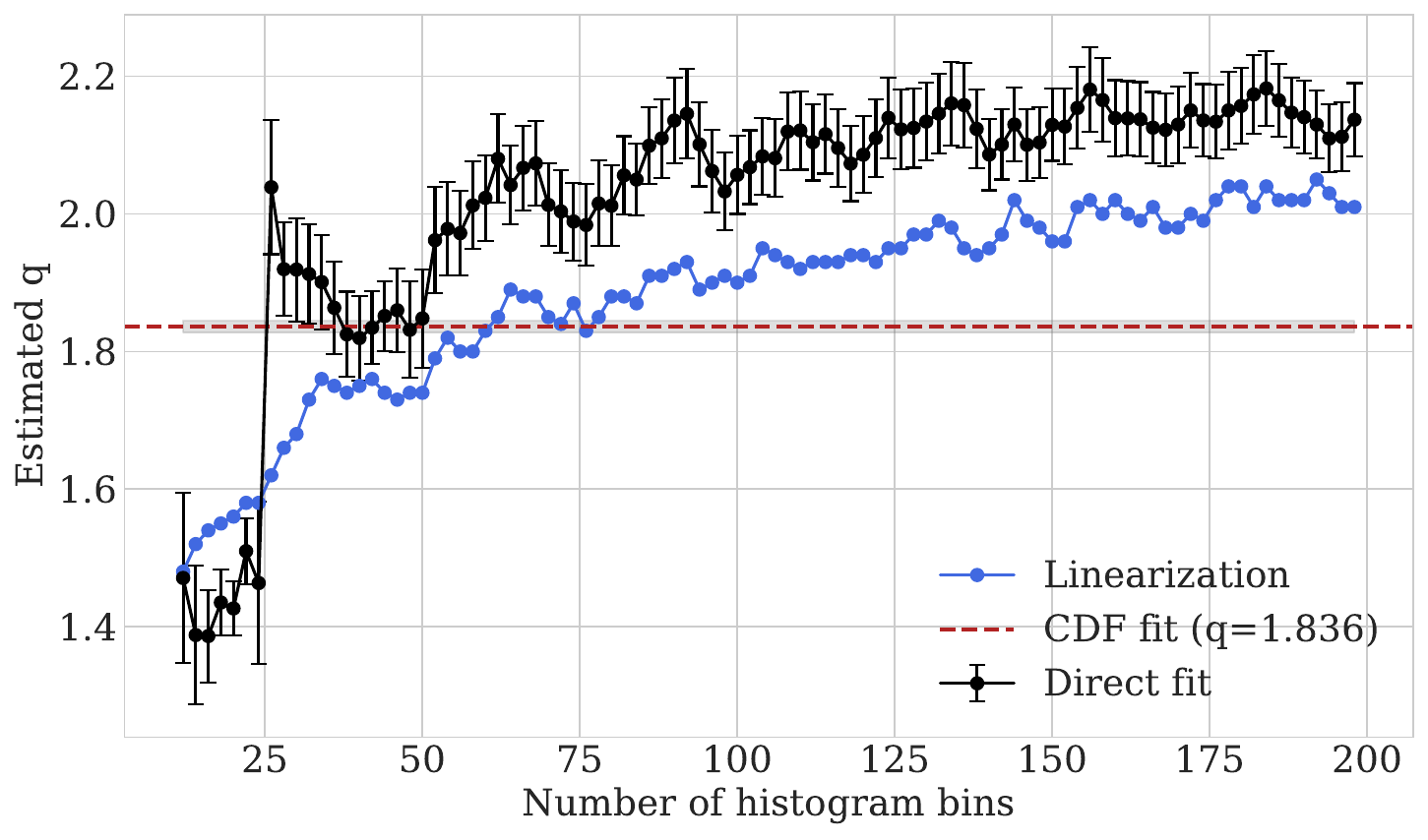}
    \caption{The parameter q, for the set of logarithmic returns of the closing value of Bitcoin, in terms of the number of bins obtained with: the direct fit (black dots), the linearaized fit (blue dots), and CDF (red dashed line).}
    \label{fig:placeholderasdazxae}
\end{figure}

\begin{figure}[H]
    \centering
    \includegraphics[width=0.75\linewidth]{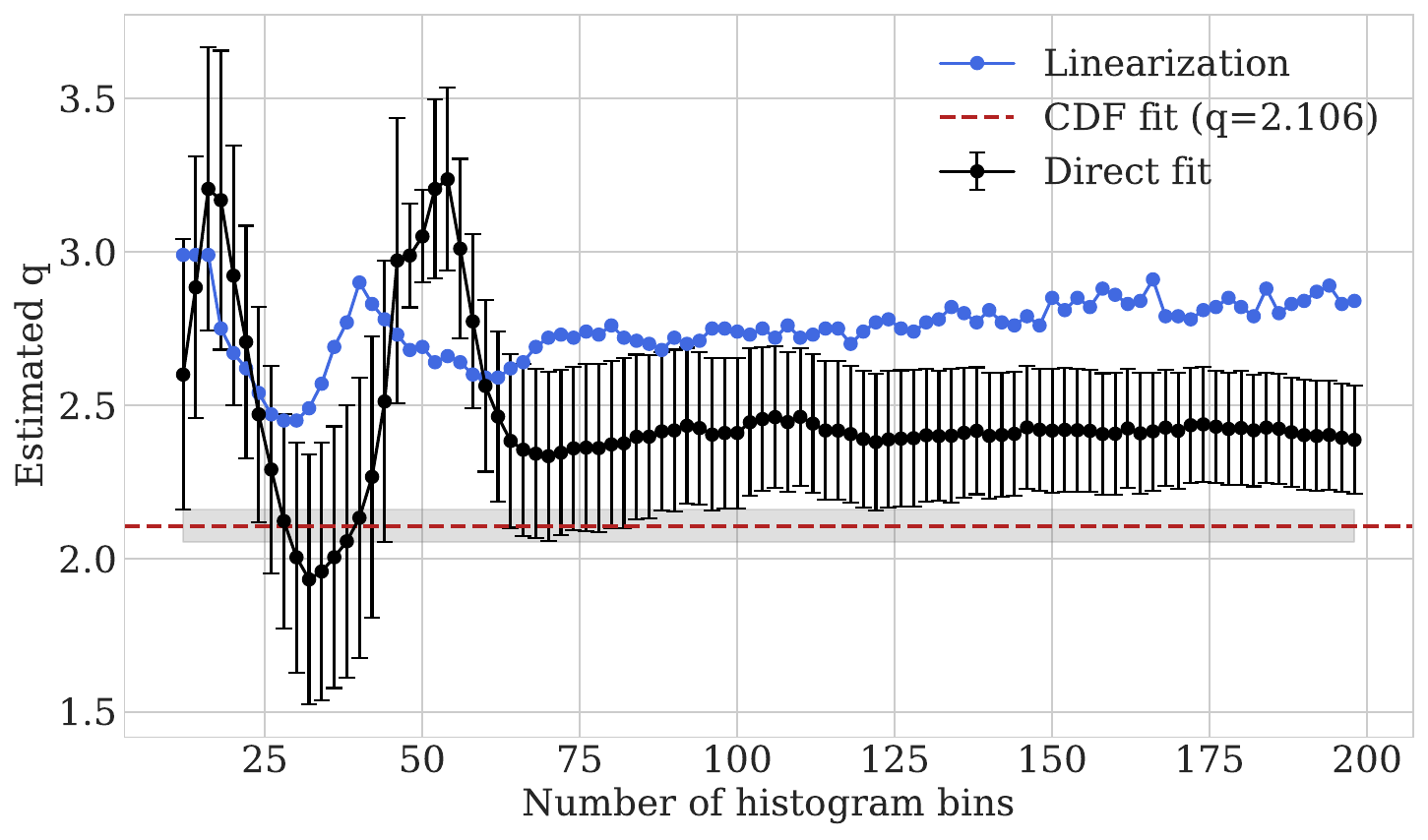}
    \caption{ The parameter q, for the set of Brazos river discharge, in terms of the number of bins obtained with: the direct fit (black dots), the linearaized fit (blue dots), and CDF (red dashed line).}
    \label{fig:placeholdeasxasr}
\end{figure}

\section{Discussion and conclusions}

Our results confirm that the observed distributions of the three systems studied are consistent with q-Gaussian heavy-tailed statistics due to constant accuracy across the range compared to standard Gaussian distributions. The fluctuations in proton density exhibit a persistent non-Gaussian character over a 45-year period, the entropic index q remains above unity, consistent with the intrinsic intermittency and long-range correlations of the solar plasma. The analysis of Bitcoin price variations highlights the extreme volatility of crypto-assets, which is notably higher than in many physical systems. This suggests that financial dynamics, driven by social and economic shocks, generate heavier tails and more frequent extreme events than those found in space plasmas. The daily flow of the Brazos River shows the highest heavy-tailed behavior among the studied cases, with q values often exceeding 2.0. This suggests that non-Gaussian models may be useful for characterizing extreme discharge fluctuations.

A key contribution of this work is the systematic comparison of three fitting techniques, which reveal significant disparities depending on the chosen approach. The direct nonlinear fitting method to the PDF, as well as fitting to linearized data, are the most widely used in the literature because they are intuitive, graphically representative of extreme events, and offer ease of interpretation of the resulting parameters. However, they appear to be sensitive to the number of bins used, since grouping events introduces noise and bias into the fit, even more so in the tails of the distributions, where events are less frequent and are the area of greatest interest in this type of analysis.

Figures 7, 8, and 9 show how the q changes with the number of bins chosen. Although the noise decreases as the number of bins increases, the difference in the results obtained is not negligible, making the subjective choice of bin number relevant. This problem cannot be solved by proposing a standard number of bins, since each dataset has a different length, and therefore it is possible to have very little data in each bin if the dataset is not large enough. As in the case of the solar wind proton density, because the data are hourly, it is possible to work with many more bins than in the other two systems, and therefore the noise is reduced and more stable results are obtained.

On the other hand, the numerical fitting method for the cumulative distribution function is an approach that avoids arbitrarily grouping data into bins and uses all the information in the dataset, accumulating it sequentially, thus eliminating bias induced by noise in the tails. The main difficulties of this method lie in its mathematical complexity, both in the interpretation of the obtained parameters and, primarily, in solving the q-Gaussian integral, which requires the use of the incomplete beta function.

Finally, we can observe that when working with a large number of bins, the first two methods begin to approach a fixed value, systematically higher than that obtained by the cumulative method. Taking all this into account, the CDF-based approach seems to provide a more stable estimate in the current datasets, although further statistical tests are required to fully establish its superiority..

\section{Data availability}

This study analyzed publicly available datasets. The data can be accessed at https://omniweb.gsfc.nasa.gov, https://coinmarketcap.com/currencies/ and https://waterdata.usgs.gov/monitoring-location/USGS-08110200. Key references describing the datasets are cited in the text. Additional results and analyses not included in the manuscript, as well as the code used, are available upon request.

\section*{Acknowledgments}
We thank GSFC/SPDF and OMNIWeb for providing access to the data. This work was financially supported by CONICET (Argentina).
\bibliographystyle{apalike}
\bibliography{atmosfera}

\end{document}